\documentstyle[aps,prl,twocolumn,epsf,float]{revtex}
\begin{document}

\twocolumn[\hsize\textwidth\columnwidth\hsize\csname @twocolumnfalse\endcsname

\title{
Collective Excitations in High-Temperature Superconductors
}

\author{
M.I.~Salkola and J.R.~Schrieffer
}
\address{
NHMFL and Department of Physics, Florida State University, 
Tallahassee, Florida 32310}

\date{December 3, 1997}
\maketitle

\begin{abstract}

Collective, low-energy excitations in quasi-two-dimensional $d$-wave
superconductors are analyzed. While the long-range Coulomb interaction
shifts the charge-density-wave and phase modes up to the plasma energy, 
the spin-density-wave excitation that arises due to a strong local
electron-electron repulsion can propagate as a damped collective 
mode within the superconducting energy gap. It is suggested that
these excitations are relevant to high-$T_c$ superconductors, close to 
the antiferromagnetic phase boundary, and may explain some of the
exotic features of the experimentally observed spectral-density and
neutron-scattering data.

\end{abstract}

\

\

]

An important aspect of superconductivity is concerned with collective
modes as they may modify low-energy properties of superconductors 
\cite{hist}. In principle, these excitations are either attributed to 
a symmetry transformation under which the system is invariant or to 
a spontaneously broken continuous symmetry. One might then expect that,
in addition to the usual charge and spin fluctuations, collective modes 
associated with the phase and the amplitude of the superconducting 
order parameter are important. Yet, in conventional superconductors 
with a momentum-independent energy gap, collective modes have essentially 
no practical significance regarding the low-energy properties of the
superconductors mainly because the long-range Coulomb interaction 
causes the phase mode to appear at the plasma energy \cite{anderson}.
The only mode that is not diffusive at long wavelengths is the amplitude 
mode, but it has no direct coupling to charge or spin degrees of 
freedom making its observation demanding \cite{lv}. 

Moreover, excitonic states that are bound pairs of quasiparticles 
have been predicted to exist in superconductors \cite{bs}. Such states
appear in angular-momentum channels other than the one in which
the Cooper pairing occurs. Theoretically, they should be present
once the effective electron-electron interaction has an attractive 
partial-wave component with a given angular momentum. Nonetheless,
there is no experimental evidence for these kind of states possibly
because of their small binding energy \cite{martin}.

While for most phenomena collective excitations in conventional 
superconductors can be ignored --- for instance, the superconducting 
energy gap in the electronic spectrum at the Fermi energy is not affected 
by these modes --- there are both theoretical and experimental reasons 
to expect that high-$T_c$ superconductors may behave differently in this 
respect. Theoretically, the energy gap has a strong momentum dependence 
and there is a large local electron-electron repulsion which may qualitatively 
change the nature of collective excitations and allow new ones to 
develop that are not related to the broken gauge symmetry. 
Experimentally, neutron-scattering studies \cite{neutron} in superconducting 
La$_{1.6-x}$Nd$_{0.4}$Sr$_x$CuO$_4$ have established the existence 
of elastic peaks in the magnetic structure factor at wave vectors 
$({1\over 2}\pm \epsilon,{1\over 2})$ and 
$({1\over 2},{1\over 2}\pm \epsilon)$ (measured in units of $2\pi/a$), 
providing direct evidence for (diffusive) spin-density-wave fluctuations
in high-$T_c$ superconducting materials. These excitations can be regarded as a
manifestation of fluctuating charge stripes and antiphase spin domains 
which have shown to provide a natural explanation
for the unusual features observed in angle-resolved photoemission 
experiments \cite{SEK}. We may also argue that they account 
for the broad ``bosonic'' feature in the electronic spectral density near 
the $({1\over 2},0)$ and $(0,{1\over 2})$ points seen only in 
the underdoped materials \cite{SS}.

In this Note, we examine under what conditions collective excitations develop
in $d$-wave superconductors and what are their experimental implications,
when the quasiparticle picture is appropriate.
Our most important observation is that $d$-wave superconductors
close to the antiferromagnetic phase boundary support a low-energy
spin-density-wave mode. As a consequence of the superconducting energy gap,
this mode can propagate coherently with reduced damping, unlike its precursor 
in the normal state where only a diffusive spin-density wave is realized. 
It can be excited by magnetic processes making it observable, for instance, 
by inelastic neutron scattering. The mode is ``massive'' because rotational 
symmetry in spin space is not broken. Under specific conditions, however, 
the mass of the mode may vanish and may even become negative signaling 
an instability of the superconducting ground state against a spontaneous 
creation of a spin-density-wave state.

Consider, for example, the Hamiltonian
\begin{equation}
H= - \sum_{ {\bf rr}' \sigma} t_{{\bf rr}'} \psi^\dagger_{{\bf r}\sigma}
\psi_{{\bf r}'\sigma}
   +\mbox{$1\over 2$} \sum_{{\bf  rr}'}
 v({\bf r}-{\bf r}') n_{\bf r} n_{{\bf r}'} - \mu\sum_{\bf r} n_{\bf r}, \label{eq:H}
\end{equation}
where $n_{\bf r}= \sum_\sigma \psi^\dagger_{{\bf r}\sigma}\psi_{{\bf r}\sigma}$ is 
the electron number operator at site ${\bf r}$, $t_{{\bf rr}'}$ is the tunneling-matrix
element between sites ${\bf r}$ and ${\bf r}'$, $\mu$ is the chemical
potential, and $v({\bf r})$ is an instantaneous 
electron-electron interaction. 
\begin{mathletters}
In describing superconducting order, it is useful to express the Hamiltonian in the form
$H= H_{\rm BCS} + H_{\rm int}$, where the BCS and interaction Hamiltonians are
\begin{eqnarray}
H_{\rm BCS} &=& \sum_{\bf k} \Psi^\dagger_{\bf k}(\epsilon_{\bf k}\hat{\tau}_3 - \Delta_{\bf k}\hat{\tau}_1)\Psi_{\bf k} \\
H_{\rm int} &=& {1\over 2N} \sum_{{\bf k}{\bf k}'{\bf q}} v({\bf q}) (\Psi^\dagger_{{\bf k}+{\bf q}}\hat{\tau}_3\Psi_{\bf k})
(\Psi^\dagger_{{\bf k}'-{\bf q}}\hat{\tau}_3\Psi_{{\bf k}'}) \\
& & \hspace*{1truecm} \mbox{ } + \sum_{\bf k} \Delta_{\bf k}\Psi^\dagger_{\bf k}\hat{\tau}_1\Psi_{\bf k}. \nonumber
\end{eqnarray}
Here, \end{mathletters}
$\Psi_{\bf k} = (\psi_{{\bf k}\uparrow} \,\, \psi^\dagger_{-{\bf k}\downarrow})^T$ is the Gor'kov-Nambu spinor, $\epsilon_{\bf k}= \sum_{\bf r} t_{{\bf r}0} e^{-{\bf q}\cdot {\bf r}}-\mu$
is the single-particle energy relative to the chemical potential, and $v({\bf q})= \sum_{\bf r} v({\bf r}) e^{-{\bf q}\cdot {\bf r}}$. The fermion operators in real and momentum spaces are related by the unitary transformation
$\psi_{{\bf r}\sigma}= N^{-1/2}\sum_{\bf k} \psi_{{\bf k}\sigma} e^{i{\bf k}\cdot {\bf r}}$, where
$N$ is the number of sites in the system.
Also, suppose that there exists 
a wave vector ${\bf Q}$ such that $\epsilon_{{\bf k}+{\bf Q}} \simeq -\epsilon_{\bf k}$; {\it i.e.}, the Fermi surface is approximately nested.
This kind of situation may qualitatively arise in lightly doped high-$T_c$ superconductors. For illustrative purposes, let only the
nearest-neighbor tunneling matrix element be non-zero so that 
$\epsilon_{\bf k}= -{1\over 2}W(\cos k_xa + \cos k_ya)-\mu$, where $W$ is 
the half bandwidth and $a$ is the lattice
spacing. Therefore, at half filling ($\mu=0$), 
${\bf Q}={\bf Q}_0$, where ${\bf Q}_0 \equiv (\pi/a,\pi/a)$. 

The energy gap $\Delta_{\bf k}$ is determined by requiring that the interaction Hamiltonian
does not give any self-energy corrections to the energy gap \cite{bob}. This condition leads 
to the gap equation, 
\begin{equation}
\Delta_{\bf k} = -{i\over 2} \sum_p v({\bf k}-{\bf p})\, {\rm Tr}\, 
\hat{\tau}_1\hat{G}({\bf p},\omega), \label{eq:gap1}
\end{equation}
where $\hat{G}({\bf p},\omega) = 1/(\omega\hat{\tau}_0 - \epsilon_{\bf p}\hat{\tau}_3 
+ \Delta_{\bf p}\hat{\tau}_1)$ denotes the Green's function of the BCS Hamiltonian. 
We use the notation in which $\sum_p=N^{-1}\sum_{\bf p}\int  (d\omega/2\pi)$ and 
$p= ({\bf p},\omega)$.

In order to determine whether the system can support collective excitations, consider an effective
two-particle interaction $\hat{\Gamma}$ that describes mutual scattering of two
quasiparticles. The poles of the effective interaction then yield the energy and lifetime of
two-particle collective excitations. In the ladder approximation, the Bethe-Salpeter equation 
for $\hat{\Gamma}$ may be written formally as
\begin{figure}
\epsfysize=1.0cm
\epsfxsize=7.5cm
\ \ \epsffile{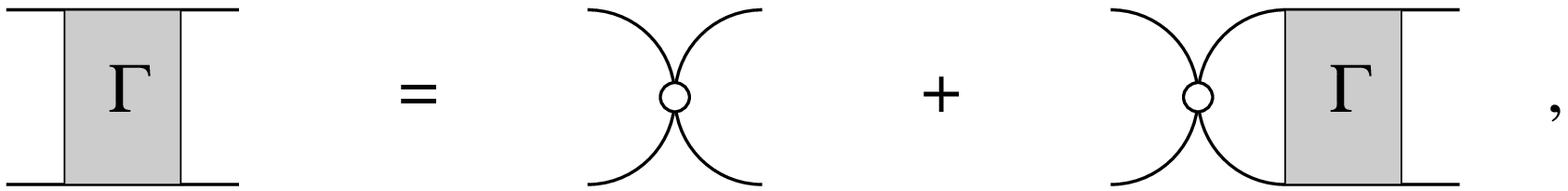}
\end{figure}
\noindent which is equivalent to the equation
\begin{equation}
\hat{\Gamma}({ k},{ k}';q)= \hat{\Gamma}_0({ k},{ k}';q) + i \sum_p 
\hat{\Gamma}_0({ k},p;q) \hat{\Lambda}(p,q)\hat{\Gamma}({ p},{ k}';q),
\end{equation}
\noindent with $\hat{\Lambda}(p,q)= \hat{G}(p+q/2)\otimes\hat{G}(p-q/2)^T$; 
here, $q$ is the total four-momentum of a quasiparticle pair.
The bare interaction vertex, denoted as
\begin{figure}
\epsfxsize=7.0cm
\ \ \epsffile{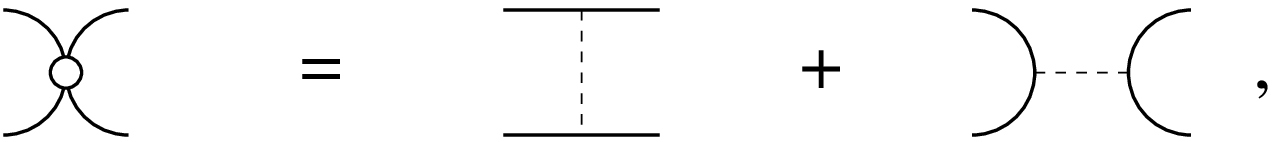}
\end{figure}
\noindent is 
 $\hat{\Gamma}_0({ k,k';q})= v({\bf k}-{\bf k}')\hat{\tau}_3 \otimes \hat{\tau}_3 - v({\bf q}) \hat{P}$,
with $\hat{P}= (\hat{\tau}_0-\hat{\tau}_1)\oplus 0$. Moreover, the outer product, $[A\otimes B]_{ab}
=A_{ij}B_{kl}$, where $a=(ij)$ and $b=(kl)$, is defined so that there is a one-to-one
correspondence between the ordered index lists $a,b \in \{1,2,3,4\}$ and $(ij),(kl)\in \{(11),(22),(12),(21)\}$.

On a square lattice, it is convenient to define orthogonal functions
$\eta_\alpha({\bf k})$ in terms of which the vertex functions and the
electron-electron interaction can
be expanded \cite{etas}. For example, $v({\bf k}-{\bf k}')= U + V_1[\cos(k_x-k'_x)a + \cos(k_y-k'_y)a]$ may be
written as $v({\bf k}-{\bf k}')= \sum_\alpha v_\alpha \eta_\alpha({\bf k}) \eta_\alpha({\bf k}')$,
where, for $\alpha=0$, $v_\alpha= U$ is the on-site electron-electron
interaction and, for $\alpha=1,\ldots,4$, $v_\alpha= V_1/2$
is the nearest-neighbor electron-electron interaction. Below, we will assume
that the on-site interaction is repulsive, $U>0$, and the nearest-neighbor interaction
is attractive, $V_1<0$, as should be appropriate for a phenomenological model
describing high-$T_c$ superconductors.
Similarly, 
$\hat{\Gamma}(k,k';q)= \sum_{\alpha\beta}
\hat{\Gamma}_{\alpha\beta}(q)\eta_\alpha({\bf k}) \eta_\beta({\bf k}')$ and
$\hat{\Gamma}_0(k,k';p)= \sum_{\alpha\beta}
\hat{\Gamma}^{(0)}_{\alpha\beta}(q) \eta_\alpha({\bf k}) \eta_\beta({\bf k}')$, with
$\hat{\Gamma}^{(0)}_{\alpha\beta}(q)= v_\alpha\delta_{\alpha\beta}\hat{\tau}_3 \otimes \hat{\tau}_3
-v({\bf q})\delta_{\alpha0}\delta_{\beta0}\hat{P}$. 
Thus, the Bethe-Salpeter equation becomes
\begin{equation}
\hat{\Gamma}_{\alpha\beta}(q) = \hat{\Gamma}^{(0)}_{\alpha\beta}(q)\
+ \sum_{\gamma\gamma'} \hat{\Gamma}^{(0)}_{\alpha\gamma}(q)  \hat{\Lambda}_{\gamma\gamma'}(q)
\hat{\Gamma}_{\gamma'\beta}(q),
\end{equation}
where $\hat{\Lambda}_{\alpha\beta}(q)= i\sum_{p}
\eta_\alpha({\bf p})\hat{\Lambda}(p,q) \eta_\beta({\bf p})$. To compactify
the notation, define new matrices such that 
$[\hat{\Gamma}]_{\alpha\beta} = \hat{\Gamma}_{\alpha\beta}$, etc.
Then, it is immediately clear that the effective interaction vertex is
\begin{equation}
\hat{\Gamma}(q)= [1-\hat{\Gamma}^{(0)}(q) \hat{\Lambda}(q)]^{-1}\hat{\Gamma}^{(0)}(q), \label{eq:sol}
\end{equation}
where $q= ({\bf q},\Omega+i0^+)$.
Because, for ${\bf q}=0$, $\hat{\Lambda}(-p,q)=\hat{\Lambda}(p,q)$, there is no mixing 
between even and odd parity sectors of $\hat{\Gamma}_{\alpha\beta}(q)$: 
the subspaces $\{\eta^{(0)}_{\bf k},\eta^{(\pm)}_{\bf k}\}$ and 
$\{\zeta^{(\pm)}_{\bf k}\}$ describing scattering in the singlet and triplet channels 
decouple. In contrast, for ${\bf q}={\bf Q}$, the subspaces 
$\{\eta^{(0)}_{\bf k},\zeta^{(\pm)}_{\bf k}\}$ and $\{\eta^{(\pm)}_{\bf k}\}$ are decoupled.
When the system has particle-hole symmetry at the Fermi energy and 
$\Omega \sim 0$, a further factorization can be shown to occur; namely, 
particle-hole and particle-particle channels decouple. This means that, for 
${\bf q}={\bf Q}$, the subspace $\{\zeta^{(-)}_{\bf k}\}$ decouples from 
$\{\eta^{(0)}_{\bf k},\zeta^{(+)}_{\bf k}\}$ and, for ${\bf q}=0$, the same occurs 
for $\{\eta^{(-)}_{\bf k}\}$ and $\{\eta^{(0)}_{\bf k},\eta^{(+)}_{\bf k}\}$. 
This is particularly convenient because
$\eta^{(-)}_{\bf k}$ determines the $d$-wave gap function, $\Delta_{\bf k}=
(\Delta_0/2)\eta^{(-)}_{\bf k}$. At zero temperature, Eq.~(\ref{eq:gap1}) becomes
$
1 = -({V_1/ N})\sum_{\bf k} {[\eta^{(-)}_{\bf k}]^2/ 4E_{\bf k}}$.
Here, $E_{\bf k}= \sqrt{\epsilon_{\bf k}^2 + \Delta_{\bf k}^2}$ is the quasiparticle
energy in the superconductor. 
We find $\Delta_0 \simeq 4W e^{-1/N_F|V_1|}$, where $W$ is the half bandwidth 
and $N_F=N(\mu)$ is the density of states at the Fermi energy in the normal state.

The collective excitations are described by the poles of the effective 
two-particle interaction, Eq.~(\ref{eq:sol}), and they are conveniently 
classified by the symmetry properties of the system.
The Hamiltonian (2) possesses a number of symmetries which lead 
to conserved currents. First, gauge symmetry, generated by the transformation
$\Psi_{\bf k}\rightarrow \Psi'_{\bf k} = e^{i\varphi\hat{\tau}_3}\Psi_{\bf k}$, 
yields charge conservation. On general grounds, one then expects that there is 
a collective mode associated with the phase of the superconducting order parameter. 
However, due to the long-range nature of the Coulomb interaction, the Anderson-Higgs 
mechanism shifts it at small momenta to the plasma energy. Second,  the symmetry 
transformation, $\Psi_{\bf k}\rightarrow \Psi'_{\bf k} = e^{\varphi\hat{\tau}_1}\Psi_{\bf k}$, 
is associated with an amplitude mode of the order parameter \cite{nambu}.
In contrast to $s$-wave superconductors, this mode is 
always over-damped in $d$-wave superconductors.  Third, spin-rotational 
symmetry leads to a new mode in the particle-hole sector, 
which is driven by the on-site Coulomb repulsion. As an example, consider
a cylindrical Fermi surface and a wave vector ${\bf q}={\bf Q}$ nesting two given ${\bf k}$
points with vanishing quasiparticle energies. For $U\sim |V_1|$, the energy $\Omega_{\bf Q}$ of 
the collective excitation is
\begin{equation}
\beta \left({\Omega_{\bf Q}\over 2\Delta_0}\right) \simeq {1\over 4}\left( {V_0\over U} - 1\right)
                   - \kappa_{\bf Q}, \label{eq:res}
\end{equation} 
when $\beta>0$. For $\Delta_0,|\mu| \ll W$, $V_0^{-1}\simeq 
{1\over N}\sum_{\bf k}{1\over 2E_{\bf k}} \sim |V_1|^{-1}$. 
The parameter $\beta=-\big[\Delta_0{\partial \log N(\epsilon)\over 
\partial \epsilon}\big]_{\epsilon=\mu}$ 
measures the magnitude of the particle-hole symmetry breaking in the density of
states $N(\epsilon)$ at the Fermi energy and ${\kappa}_{\bf Q}$ describes
the contribution due to the mixing between the particle-hole and particle-particle channels.
For $U,|V_1|\ll W$, it 
can be expanded as ${\kappa}_{\bf Q} = {\kappa}_{\bf Q}^{(1)} + {\kappa}_{\bf Q}^{(2)}$, 
where ${\kappa}_{\bf Q}^{(1)} \sim 2\big({V_1\over W}\big)^2\big( {\beta\over 2\pi}\big)^2 $
and ${\kappa}_{\bf Q}^{(2)} \sim  \big({V_1\over W}\big)^2
\big({\Omega_{\bf Q}\over \pi\Delta_0}\big)^2$. For $\beta < 0$, the collective
excitation remains massive ($\Omega_{\bf Q}>0$) for all values of $V_0/U$, 
down to the point where the superconductor becomes unstable.
The excitation is damped, because it overlaps with the quasiparticle
continuum. For the other values of ${\bf q}$, the continuum
does not necessarily start from zero but at some finite value $\vartheta_{\bf q}
\equiv \min_{\bf k}(E_{\bf k} + E_{{\bf k}+{\bf q}})$. It is then
possible to have collective excitations with an infinite lifetime,
when $\Omega_{\bf q} < \vartheta_{\bf q}$.

The exact nature of the collective excitation is deduced by examining its coupling
to the spin operator $S_z({\bf q}) = 
\sum_{\bf k} \Psi^\dagger_{{\bf k}}\hat{\gamma}^{(0)}_z\Psi_{{\bf k}+{\bf q}}$,
where $\hat{\gamma}^{(0)}_z= \mbox{$1\over 2$} \hat{\tau}_0$. That the mode 
is a spin-density wave becomes evident by computing the spin correlation function, 
$\chi({\bf q},\tau)= \langle T_\tau S_z({\bf q},\tau)S_z(-{\bf q},0)\rangle$, 
\begin{figure}
\epsfxsize=7.5cm
\ \ \epsffile{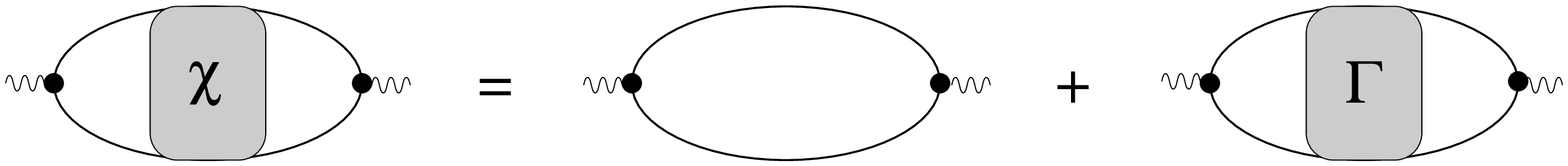}
\end{figure}
\noindent Specifically, its temporal Fourier transform is given by the formula
\begin{equation}
\chi(q)= \langle \gamma_z|\hat{\Lambda}(q)[1 + \hat{\Gamma}(q)\hat{\Lambda}(q)]
|\gamma_z\rangle, \label{eq:sus}
\end{equation}
where the column vector $|\gamma_z\rangle$ is defined as 
$\langle a\alpha|\gamma_z \rangle= {1\over 2}[\hat{\tau}_0]_{ij}\delta_{\alpha 0}$. 
In the limit $\Omega\rightarrow 0$, our result reduces to the form obtained
by non-conserving approximation \cite{ns}, when the energy spectrum
has particle-hole symmetry at the Fermi energy. However, for energies 
$\Omega \gtrsim \Delta_0$ or in the absence of particle-hole symmetry,
one must use the general result, Eq.~(\ref{eq:sus}).
Note that, by incorporating the mixing of the particle-hole and
particle-particle degrees of freedom, $\kappa_{\bf Q}$ accounts for, for example, 
the effect of any two-particle, spin-triplet excitations \cite{dz} at
the momentum ${\bf Q}$. Similar calculation shows that 
no resonance develops for the charge response near  ${\bf q}={\bf Q}$. 

The most favorable 
conditions for observing these excitations are most likely found 
in underdoped high-$T_c$ superconductors close to the antiferromagnetic 
phase boundary. In the antiferromagnetic phase, the above collective 
mode is replaced by the Goldstone mode of the antiferromagnet.
Interestingly, for $U \gtrsim V_0$, the system is unstable against
a spontaneous creation of quasiparticle-quasihole virtual bound 
states at ${\bf q}={\bf Q}$. This implies a phase transition to 
an antiferromagnetic state. In contrast to $d$-wave superconductors, 
the Ward identity \cite{lv,nambu} excludes the spin-density-wave 
collective mode in conventional superconductors with a momentum 
independent gap function. 

\begin{floating}[t]
\narrowtext
\begin{figure}
\setlength{\unitlength}{1truecm}
\begin{picture}(8.5,4.6)
\put(0.0,-5.6){\epsfbox{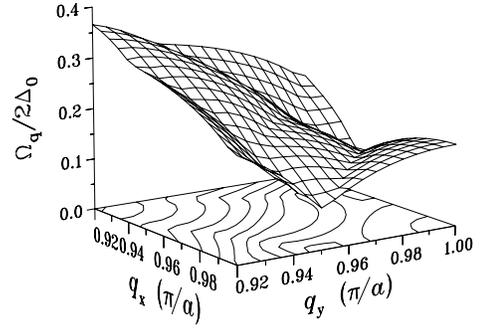}}
\end{picture}
\caption[]{\small
\noindent The energy $\Omega_{\bf q}$ of the spin-density-wave excitation 
as a function of wave vector ${\bf q}$ at zero temperature for $U/W= 0.485$ 
and $V_1/W=-0.608$. Here, the superconducting energy gap is $\Delta_0/W = 0.1$ 
and the chemical potential is chosen so that the density of holes equals 10\% 
(relative to half filling). The excitation energy $\Omega_{\bf q}$ is
computed numerically from the exact result, Eq.~(\ref{eq:sus}).
}
\end{figure}
\vskip -.5truecm\end{floating}

To obtain a quantitative understanding of the dispersion relation of the
collective mode, we resort to numerical methods. Figure 1 shows the energy
of the spin-density wave as a function of wave vector near half filling with
hole density equal to 10\%. The minimum value of the excitation energy,
$\Omega_{\bf q}/2\Delta_0 \simeq 0.08$, is located at the wave vector 
${\bf q}\simeq(0.95\pi/a,\pi/a)$, implying that the superconductor 
becomes unstable against a spontaneous creation of an antiferromagnetically 
ordered state with vertical (horizontal) antiphase domain walls. 
In contrast to neutron scattering data \cite{neutron} in superconducting
La$_{1.6-x}$Nd$_{0.4}$Sr$_x$CuO$_4$ which show that the
mean separation $\ell$ between antiphase domain walls should scale with 
the hole density $x$ as $\ell \simeq a/2x$, we find that $\ell$ is more than 
by a factor of two longer than the experimental one. (In the present 
approximation, $\ell$ also depends on $\Delta_0$.)
The failure to predict correctly the domain-wall periodicity
is similar to the problem of describing the static, incommensurate 
stripe order in the Hartree-Fock approximation \cite{schulz}. 
Although at zero temperature exciting vertical spin-density-wave 
fluctuations requires the least amount of energy, the rotation of their 
orientation relative to the underlying lattice constitutes a relatively 
soft mode; see Fig.~1. For example, 
the excitation energy at the saddle point ${\bf q}=0.98{\bf Q}_0$ is 
about 60\% larger than the minimum energy required to excite 
a vertical mode. The form of the dispersion relation $\Omega_{\bf q}$
is affected by the lifetime effects: the troughs clearly visible
in Fig.~1 mark the boundaries between damped ($\Omega_{\bf q} > \vartheta_{\bf q}$)
and undamped ($\Omega_{\bf q} < \vartheta_{\bf q}$) excitations. 
For example, the excitations with ${\bf q}={\bf Q}_0$ and $0.94{\bf Q}_0$ 
are undamped whereas the excitation with ${\bf q}=0.97{\bf Q}_0$ is damped.

\begin{floating}[t]
\narrowtext
\begin{figure}
\setlength{\unitlength}{1truecm}
\begin{picture}(8.5,9.7)
\put(0.0,-0.7){\epsfbox{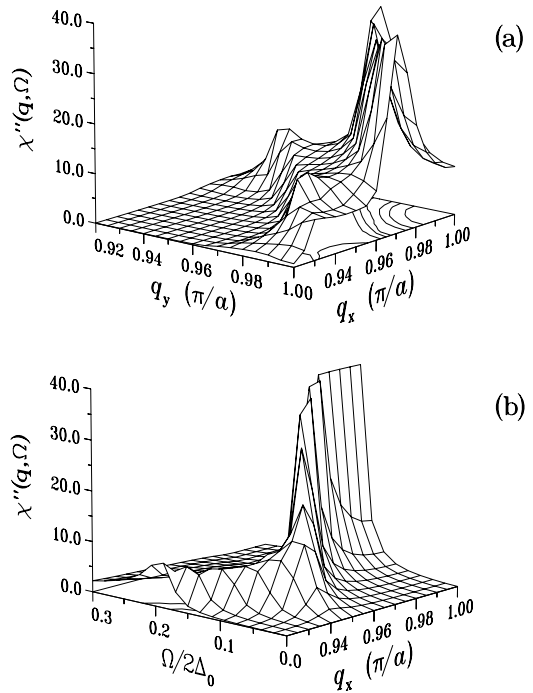}}
\end{picture}
\caption[]{\small
\noindent The imaginary part of the spin correlation function 
$\chi''({\bf q},\Omega)$ as a function of (a) the wave vector ${\bf q}$ 
with $\Omega/2\Delta_0=0.13$ and (b) $\Omega$ and $q_x$ with $q_y=\pi/a$
at zero temperature for $U/W= 0.485$ and $V_1/W=-0.608$. 
A small broadening of resonances is obtained by surmising a finite energy 
resolution of magnitude $\gamma/W=3\cdot 10^{-3}$ due to quasiparticle
lifetime effects. Here, the superconducting energy gap is $\Delta_0/W = 0.1$ 
and the chemical potential is chosen so that the density of 
holes equals 10\% (relative to half filling).
}
\end{figure}
\vskip -.5truecm\end{floating}

Figure 2 illustrates the behavior of the imaginary part of the spin
correlation function, Eq.~(\ref{eq:sus}). The collective excitation 
produces a distinctive resonance structure in $\chi''({\bf q},\Omega)$ 
at energies specified by the dispersion relation, $\Omega= \Omega_{\bf q}$. 
At low energies, the resonance becomes narrower as the excitation energy 
$\Omega_{\bf q}$ decreases. However, in the vicinity of the $(\pi/a,\pi/a)$ 
point, for example, the collective excitation appears below the onset energy 
of the two-particle continuum $\vartheta_{\bf q}$. Accordingly, the collective
excitation would acquire an infinite lifetime yielding a resolution-limited 
peak in $\chi''({\bf q},\Omega)$ --- also, below this onset energy, the usual 
quasiparticle contribution to $\chi''({\bf q},\Omega)$ would vanish --- 
if the lifetime of quasiparticles were infinite. Clearly the collective 
mode is an important new feature describing the low-energy spin correlation 
function \cite{note}. 

In a $d$-wave superconductor, strongly-scattering impurities induce virtual 
bound states \cite{us2} by modifying the local potential energy of 
electrons at the impurity site. In addition, they may also change local 
Coulomb interaction between electrons at the impurity site. Such an effect 
can be accounted for by including the term $U_{\rm imp} 
n_{{\bf r}_0\uparrow}n_{{\bf r}_0\downarrow}$, where ${\bf r}={\bf r}_0$ 
is the location of the impurity. A straightforward calculation shows
that this effect leads to a virtual bound state with the energy
$\Omega_0/2\Delta_0 \sim 2\sqrt{U_c/U_{\rm imp} - 1}$, where 
$U_c \sim 1/\pi N_F$.  In contrast to an impurity potential, 
the impurity interaction has a critical value $U_c$, below which 
the ground state is nonmagnetic and the spin quantum number of 
the resonance state equals to ${1\over 2}$, and above which the 
impurity becomes magnetic in the sense that one electron spin is 
trapped to the impurity site.

Finally, one may ask whether excitonic states of bound pairs of quasiparticles
are feasible in $d$-wave superconductors. It is immediately clear that while
the effective interaction has an attractive partial wave in the (extended) 
s-wave channel, it is nevertheless too weak near half filling to
support any excitons with the same angular quantum number. 
It is only far away from half filling that these states might appear 
as virtual bound states because of the proximity to a superconducting 
state with extended $s$-wave symmetry.

In conclusion, we have shown that $d$-wave superconductors can support
propagating collective modes that are best described as spin-density waves. It is
then natural to anticipate that the most favorable conditions for detecting
them are found close to the antiferromagnetic phase boundary. Furthermore, 
superconducting fluctuations couple particle-hole and particle-particle excitations 
allowing the latter ones to be probed by the usual means. This coupling 
is particularly important if the two excitations are (nearly) degenerate.

We would like to thank D.Scalapino, E.Demler, A.Fetter, 
S.Rabello, and S.Zhang for useful discussions. This work was supported 
by the NSF under Grant Nos. DMR-9527035 and DMR-9629987.


\begin{references}

\bibitem{hist}  Historically, collective excitations had an important 
role in establishing the gauge-invariant theory of superconductivity; 
for example, see Ref.~\cite{bob} below.

\bibitem{anderson} P.W.Anderson, Phys. Rev. {\bf 112}, 1900 (1958).
 
\bibitem{lv} P.B.Littlewood and C.M.Varma, Phys. Rev. B{\bf 26}, 4883 (1982).

\bibitem{bs} A.Bardasis and J.R.Schrieffer, Phys. Rev. {\bf 121}, 1050 
(1961); and references therein.

\bibitem{martin} P.C.Martin, in {\it Superconductivity}, Ed. R.D.Parks
(Dekker, New York, 1969), Vol. I, 371.

\bibitem{neutron} J.M.Tranquada {\it et al}., Phys. Rev. Lett. {\bf 78}, 338 (1997).

\bibitem{SEK} M.I.Salkola, V.J.Emery, and S.A.Kivelson, Phys. Rev. Lett.
{\bf 77}, 155 (1996).

\bibitem{SS} Z.-X.Shen and J.R.Schrieffer, Phys. Rev. Lett. {\bf 78}, 
1771 (1997).

\bibitem{bob} J.R.Schrieffer, {\it Theory of Superconductivity} (Benjamin-Cummings, Reading, 1983).

\bibitem{etas} They are defined as $\eta_0({\bf k})\equiv \eta^{(0)}_{\bf k}= 1$, 
                                   $\eta_1({\bf k})\equiv \eta^{(+)}_{\bf k}= \cos k_xa + \cos k_ya$,
                                   $\eta_2({\bf k})\equiv \eta^{(-)}_{\bf k}= \cos k_xa - \cos k_ya$,
                                   $\eta_3({\bf k})\equiv \zeta^{(+)}_{\bf k}= \sin k_xa + \sin k_ya$,
                                   $\eta_4({\bf k})\equiv \zeta^{(-)}_{\bf k}= \sin k_xa - \sin k_ya$, etc.

\bibitem{nambu} Y.Nambu, Phys. Rev. {\bf 117}, 648 (1960).

\bibitem{ns} N.Bulut and D.J.Scalapino, Phys. Rev. B{\bf 53}, 5149 (1996).

\bibitem{dz} E.Demler and S.C.Zhang, Phys. Rev. Lett. {\bf 75}, 4126 (1995); 
E.Demler {\it et al.} (unpublished).

\bibitem{schulz} H.J.Schulz, Phys. Rev. Lett. {\bf 64}, 1445 (1990).

\bibitem{note} Further contact with experiments \cite{neutron} can be made by noting
that impurity moments pin the spin-density-wave mode leading to
a frozen spin order. The resulting incommensurate, elastic peaks in magnetic
scattering are expected to disappear at a temperature $T_p$, which scales with 
the binding energy and therefore decreases as $\Omega_{\bf Q}$ increases.
Thus, $T_p$ decreases with the increasing hole density.

\bibitem{us2} M.I.Salkola, A.V.Balatsky, and D.J.Scalapino, Phys. Rev. Lett. 
{\bf 77}, 1841 (1996); M.I.Salkola, A.V.Balatsky, and J.R.Schrieffer, Phys. Rev. B{\bf 55}, 
12648 (1997).

\end{references}
\end{document}